\documentclass[sigconf,nonacm]{acmart}

\usepackage{multirow}
\usepackage{tikz}

\newcommand\copyrightnotice[1]{
	\begin{tikzpicture}[remember picture,overlay]
	\node[anchor=south,yshift=10pt] at (current page.south) {\fbox{\parbox{\dimexpr\textwidth-\fboxsep-\fboxrule\relax}{#1}}};
	\end{tikzpicture}
}

\AtBeginDocument{%
  \providecommand\BibTeX{{%
    \normalfont B\kern-0.5em{\scshape i\kern-0.25em b}\kern-0.8em\TeX}}}

\copyrightyear{2020}
\acmYear{2020}
\setcopyright{acmlicensed}\acmConference[ARES 2020]{The 15th International Conference on Availability, Reliability and Security}{August 25--28, 2020}{Virtual Event, Ireland}
\acmBooktitle{The 15th International Conference on Availability, Reliability and Security (ARES 2020), August 25--28, 2020, Virtual Event, Ireland}
\acmPrice{15.00}
\acmDOI{10.1145/3407023.3407030}
\acmISBN{978-1-4503-8833-7/20/08}

\begin{document}

\title{Analyzing the Real-World Applicability of DGA Classifiers}

\author{Arthur Drichel}
\affiliation{%
	\institution{RWTH Aachen University}
}
\email{drichel@itsec.rwth-aachen.de}

\author{Ulrike Meyer}
\affiliation{%
	\institution{RWTH Aachen University}
}
\email{meyer@itsec.rwth-aachen.de}

\author{Samuel Sch{\"u}ppen}
\affiliation{%
	\institution{Siemens CERT}
}
\email{samuel.schueppen@siemens.com}

\author{Dominik Teubert}
\affiliation{%
	\institution{Siemens CERT}
}
\email{dominik.teubert@siemens.com}

\renewcommand{\shortauthors}{Drichel et al.}

\begin{abstract}
Separating benign domains from domains generated by DGAs with the help of a binary classifier is a well-studied problem for which promising performance results have been published. The corresponding multiclass task of determining the exact DGA that generated a domain enabling targeted remediation measures is less well studied. Selecting the most promising classifier for these tasks in practice raises a number of questions that have not been addressed in prior work so far. These include the questions on which traffic to train in which network and when, just as well as how to assess robustness against adversarial attacks. Moreover, it is unclear which features lead a classifier to a decision and whether the classifiers are real-time capable. In this paper, we address these issues and thus contribute to bringing DGA detection classifiers closer to practical use. In this context, we propose one novel classifier based on residual neural networks for each of the two tasks and extensively evaluate them as well as previously proposed classifiers in a unified setting. We not only evaluate their classification performance but also compare them with respect to explainability, robustness, and training and classification speed. Finally, we show that our newly proposed binary classifier generalizes well to other networks, is time-robust, and able to identify previously unknown DGAs.
\end{abstract}




\maketitle
\copyrightnotice{\copyright\space Copyright held by the owner/author(s) 2020. This is the author's version of the work. It is posted here for your personal use. Not for redistribution. The definitive version was published in Proceedings of the 15th International Conference on Availability, Reliability and Security (ARES 2020), https://doi.org/10.1145/3407023.3407030}
\section{Introduction}
\label{sec:introduction}

Bots need to be able to establish a connection to their command and control (C2) server in order to obtain updates or commands. To this end, they often rely on domain generation algorithms (DGAs), which generate a large amount of pseudo-random domain names using a seed. The botnet herder knows the seed and is thus able to predict the algorithmically generated domains (AGDs) in advance and to register a small subset of these domains. The bots query the AGDs one-by-one trying to obtain a valid IP address for the C2 server.
The majority of these queries result in non-existent domain (NXD) responses. Only the domain names that are registered by the botnet herder successfully resolve to valid IP addresses and thus to a successful contact between the bot and the C2 server.

To detect the activity of DGA-based malware, various binary machine learning classifiers have been proposed that label domain names contained in DNS queries as benign or malicious. These include classical feature-based approaches such as random forests (RFs) or support vector machines (SVMs) (e.g., \cite{schuppen_fanci_2018}), as well as featureless classifiers, such as recurrent (RNNs) or convolutional neural networks (CNNs) (e.g., \cite{woodbridge_predicting_2016, yu_character_2018, saxe_expose_2017}). The classification performance reported for these binary classifiers is so promising that it seems high time to bring these classifiers into practice. 

While binary classification allows to detect infections with DGA-based malware in a network, multiclass classification additionally allows for attributing an AGD to a specific malware family. This ultimately makes it possible to trigger appropriate remediation measures. The multiclass task is far less well studied and has proven to be more challenging than the binary task (e.g., \cite{woodbridge_predicting_2016,tran_lstm_2018,sivaguru_evaluation_2018}).   

Choosing among the available approaches and tuning them for real-world usage raises a number of questions, starting with which one of the various proposed classifiers to use for each task. Taking a closer look at previous publications shows that the proposed classifiers are currently hard to compare as there is barely any comparative evaluation of their performance, especially not in a unified setting based on the same real-world data. Thus, it is currently hard to assess the classification performance and training and classification speed of the various classifiers, let alone compare what these classifiers learn and how robust they are against evasion with the help of adversarial examples. Bringing them closer to practice also requires addressing the question in which network to train, which traffic to use, and how often to train.

In this paper, we address these issues and thus further pave the way for a smooth adoption of DGA classifiers in practice. In particular, we make the following contributions:

First, we propose two classifiers for DGA detection based on residual neural networks (ResNets), one for binary and one for multiclass classification. Our classifiers operate on domain names alone and do not need any additional contextual information. We focus on domain names extracted from non-resolving DNS traffic (NX-traffic), which has several advantages in practice. Foremost, the activity of a DGA is typically recognizable in the NX-traffic well before one of the AGDs successfully resolves. Thus, a bot's activity can be detected even before it is commanded to partake in any malicious action. In addition, the amount of non-resolving DNS traffic is an order of magnitude smaller than the amount of full DNS traffic and thus easier to monitor. 

Second, we specify a unified setting to comprehensively compare our proposed classifiers with the best feature-based approach and the previously proposed types of featureless classifiers. To guarantee a fair comparison we evaluate all classifiers on the same real-world NX-traffic in combination with malicious data obtained from DGArchive\cite{plohmann_comprehensive_2016} using the same unified input data pre-processing. 
For the binary classification task, we show that our classifier performs at least comparable to the best state-of-the-art approaches, has a very low false positive rate, and clearly exceeds the other classifiers in the extraction of complex features. For the multiclass classification task, our ResNet-based classifier outperforms prior work in attributing AGDs to DGAs accomplishing   an improvement of over 5\% in f1-score while requiring 30\% less training time compared to the next best classifier.

Third, we study how the classifiers perform with respect to explainability, robustness, generalizability, and training and classification speed. In particular, we show that the ResNet-based binary classifier generalizes well to new environments, performs stable over time, is capable of performing live detection within large networks, is robust against two recently proposed adversarial attacks, and able to detect new DGAs as well as new seeds of known DGAs in a real-world test.

\section{Related Work}
\label{sec:related_work}

Various approaches have been proposed to detect DGA activities within networks. In \cite{grill_detecting_2015} meta information of transmitted network packets are statistically analyzed to detect DGA malware infected devices. Others use filtering and clustering techniques to identify C2 servers (e.g., \cite{schiavoni_phoenix_2014,yadav_winning_2012}). Many works make use of machine learning classifiers, such as extreme learning machines, SVMs, or different types of Decision Tree algorithms, in combination with manual feature engineering to classify domain names as benign or malicious (e.g.,\cite{shi_malicious_2018, bilge_exposure_2014, schuppen_fanci_2018}). In \cite{antonakakis_throwaway_2012} a system is proposed which uses clustering combined with a hidden Markov Model to determine the likelihood that a domain is an AGD generated by a certain DGA. All of these approaches but \cite{schuppen_fanci_2018} rely on extensive tracking of DNS traffic for classification. As such, they are more intrusive and resource intensive than approaches performing classification on information extracted from a single domain. 

In this latter category, different types of deep learning based classifiers (RNNs or CNNs) have been proposed for the DGA binary (e.g., \cite{woodbridge_predicting_2016, yu_character_2018, saxe_expose_2017}) and the DGA multiclass classification task (e.g.,\cite{woodbridge_predicting_2016,tran_lstm_2018,sivaguru_evaluation_2018}). The advantage of the deep learning based approaches is that they do not require the extensive feature engineering, which is necessary for the classical ones. Different comparison papers (e.g.,\cite{yu_character_2018,sivaguru_evaluation_2018,spooren_detection_2019}) show that the deep learning classifiers outperform the classical approaches. However, the feature-based classifiers which are used as a baseline in these works are often developed with little effort or a certain input pre-processing is used which renders several features of state-of-the-art feature-based approaches useless. Thus, related work falls somewhat short of providing a fair comparison of these two types of approaches. Moreover, all prior deep learning based classifiers are evaluated on full DNS traffic or on its resolving part rather than on NX-traffic. Often, even artificial data sets are used relying on website popularity rankings such as Alexa~\cite{alexa} for benign ground truth data. A recent study \cite{lepochat_tranco_2019} shows that such website popularity rankings are easy to manipulate. An adversary could inject arbitrary malicious domain names in such lists with only little effort making these data sources untrustworthy. In our work, we overcome these shortcomings by providing a fair comparison of a state-of-the-art feature-based approach and several deep learning based approaches, including our new ResNet-based classifiers, using the same real-world NX-traffic data for all experiments and classifiers.

The deep learning classifier for the multiclass task proposed in \cite{woodbridge_predicting_2016} is prone to class imbalances. As shown in \cite{zhi-huazhou_training_2006} multiclass classification does not benefit from common approaches such as over- or undersampling. Rather, these techniques can have a negative impact on the performance of a classifier. An approach to overcome this problem is proposed in \cite{tran_lstm_2018} and involves making the classifier cost-sensitive. Thereby, the authors are able to greatly increase the detection rate of DGAs which are less represented in the training data.

Recently, some of the DGA classifiers have been shown to be vulnerable to adversarial examples which are worst-case perturbations of input data causing a classifier to misclassify such samples (e.g. \cite{spooren_detection_2019,peck_charbot_2019,anderson_deepdga_2016,sidi_maskdga_2019}). In \cite{spooren_detection_2019} the samples are generated by exploiting known features of a feature-based classifier. In \cite{peck_charbot_2019} they are created by slightly perturbing Alexa top domain names, exploiting the fact that many of the classifiers are trained using resolving DNS traffic. In \cite{anderson_deepdga_2016} a generative adversarial network is leveraged to generate adversarial samples. Finally, in \cite{sidi_maskdga_2019} perturbation is added to an AGD of a known DGA based on the computed  Jacobian-based Saliency Map for a specific input sample. The effectiveness of the adversarial samples generated by these strategies have so far only been evaluated against classifiers trained on resolving DNS traffic. We show that, when trained on NX-traffic, the classifiers are more robust and correctly classify the adversarial samples generated by the strategies described in \cite{spooren_detection_2019,peck_charbot_2019}.
\section{Unified Input Processing \& ResNet}
\label{sec:resnet}

In this section, we propose two different versions of ResNet-based classifiers, B-ResNet for DGA binary
classification and M-ResNet for DGA multiclass classification. Alongside, we introduce the unified input processing we use to comparatively assess the performance of the various available classifiers. ResNets \cite{he_deep_2016, he_identity_2016} have achieved great results in the ImageNet Large Scale Visual Recognition Challenge (ILSVRC)\cite{russakovsky_imagenet_2015}. They are built up of stacked residual building blocks, which introduce skip connections allowing the gradient to bypass layers in unchanged form during training. The effectiveness of ResNets is caused by the skip connections making them behave like ensembles of relatively shallow neural network classifiers \cite{veit_residual_2016}. Hence, making ResNets deeper correlates to the effect of increasing the number of shallow networks of an ensemble classifier.    
Our binary classification task is less complex and a shallow network is already sufficient to achieve accuracies comparable to other state-of-the-art approaches. However, for the more challenging multiclass classification task, the network we propose is deeper and able to achieve an improvement of more than 5\% in f1-score compared to the best state-of-the-art approach. 

\subsection{Unified Input Pre-Processing}
\label{subsection:pre_processing}
To enable a fair comparison of the ResNet-based classifiers with other featureless classifiers, we need to unify the input pre-process-ing. Hence, before we handover the NXDs to any of the featureless classifiers we perform a pre-processing step in which we convert all characters to lowercase and map every valid character to a unique integer. Additionally, we pad the input with zeros from the left side to the maximum domain length of 253 characters \cite{mockapetris_domain_1987} to be able to perform training and classification on every possible NXD while using batch learning. In contrast to other works \mbox{(e.g.\ \cite{woodbridge_predicting_2016, spooren_detection_2019})}, we do not remove the top-level domain (TLD) or any labels which extend past the second level domain. This information is crucial for a fair comparison with the currently best feature-based approach~\cite{schuppen_fanci_2018} as many of its features depend on it. Note that experiments in which we discard the TLD or labels past the second level showed a loss in accuracy also for the deep learning based approaches. 

\begin{figure}[!t]
	\centering
	\includegraphics[scale=0.489]{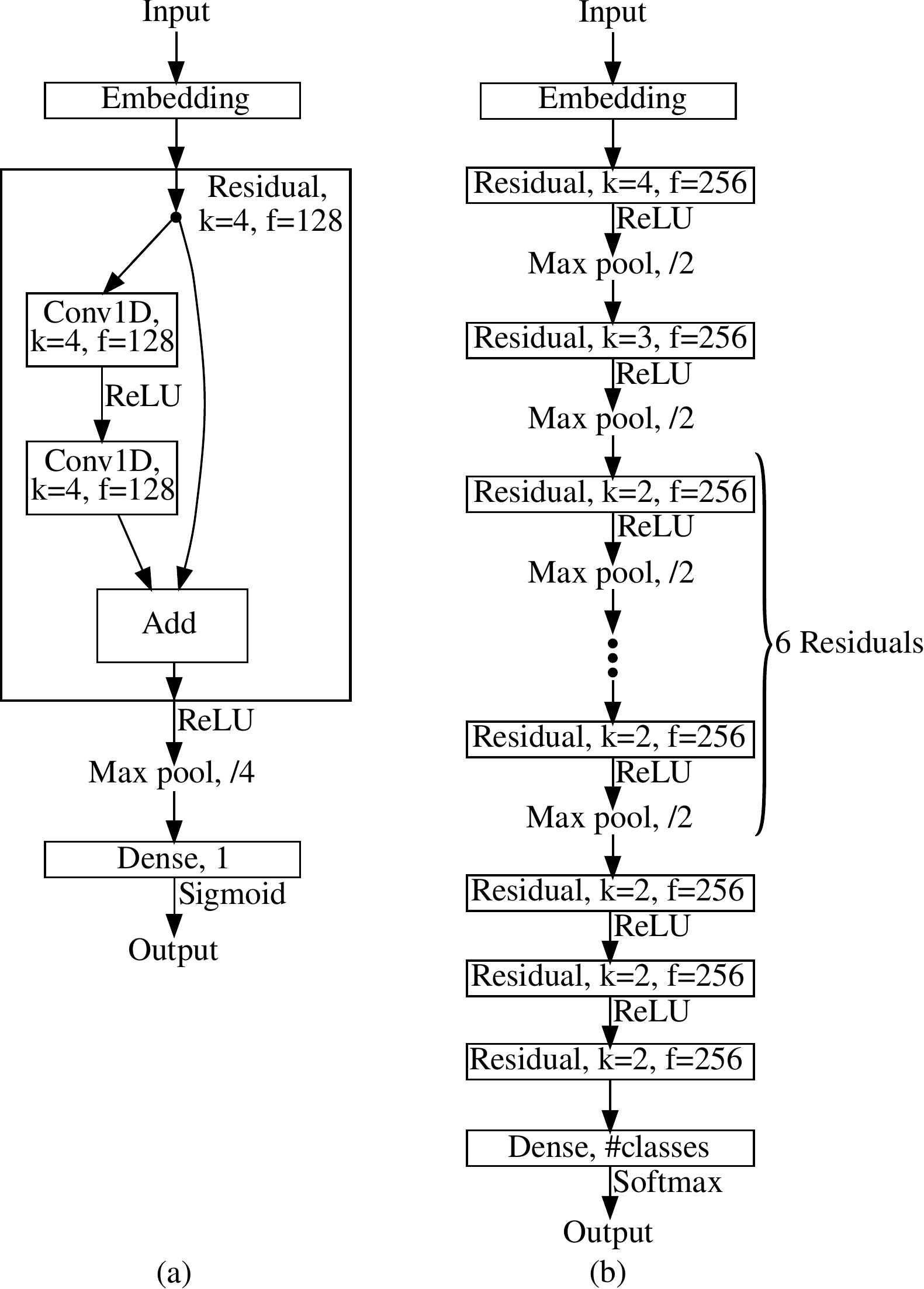}
	\caption{(a) B-ResNet network architecture with insight into a single residual block used for binary classification. (b) M-ResNet architecture used for the multiclass classification.}
	\label{fig:combined_models}
\end{figure}

\subsection{B-ResNet}
\label{subsection:b-resnet}
The network architecture for B-ResNet and the internals of one residual building block are depicted in Fig.\ref{fig:combined_models}(a). B-ResNet starts processing the input data using an embedding layer, which incorporates additional information about the relations between single characters into the encoding of a domain name. This is achieved by projecting similar characters to similar vectors where the understanding of similarity is learned automatically based on the training data. Thereby, the embedding layer learns a 128-dimensional vector representation for every valid character. Consequently, every NXD which is handed over to the classifier is projected by the embedding layer into a sequence of 128-dimensional vectors of length 253. The weights of the embedding layer are jointly optimized with all other weights of the neural network during training. For binary classification, we use a single residual block. The input to the residual block passes two convolutional layers where in between the rectified linear unit (ReLU) non-linearity function is applied to the outcome of the first convolutional layer. Additionally, the unchanged input is added to the outcome of the second convolutional layer. The residual block is specified by two variables which determine the parameters of the two convolutional layers. The first one is the kernel size $k$ and the second parameter $f$ specifies the number of filters within the convolutional layers. For binary classification, $k$ is set to $4$ and $f$ to $128$. Additionally, we use ReLU on the output of the residual block and perform max pooling with a pooling size of $4$. Lastly, the final output layer, consisting of a single node with sigmoid activation, performs the logistic regression for the binary classification. For training the classifier, we use the binary cross-entropy as loss function and Adam\cite{kingma_adam_2015} as optimization algorithm with a batch size of 128.

\subsection{M-ResNet}
\label{subsection:m-resnet}
The architecture of our M-ResNet model is depicted in Fig.\ref{fig:combined_models}(b). We use eleven residual blocks where the first two use a kernel size of $4$ and $3$, respectively. All other residual blocks are set to a kernel size of $2$. We use an embedding layer with the same output size as B-ResNet but increase the number of filters of the residual blocks to 256. Due to this change we need to increase the dimension of the input to the first residual block in order to be able to perform the addition operation. This is done by using an additional convolution with $k=1$ and $f=256$ before summing up the input to the outcome of the other branch. During our studies we experimented with different depths and obtained best results for classifiers with eleven to 14 residual blocks. We chose to further evaluate the classifier with eleven residual blocks since it requires significantly less training time than the ones with more blocks. Extending the network architecture with more than 14 residual blocks did not improve the performance of the classifier. After each residual block we apply ReLU and perform max pooling with a pooling size of $2$ until the input data dimension reduces to one single vector. Before the final layer, we do not apply ReLU on the output of the last residual block. In contrast to the binary classification, the final output layer is now composed of as many nodes as classes are present. Here, it performs a multinomial logistic regression by using the softmax activation function. For training, the categorical cross-entropy is used to compute the loss and Adam is used as optimization algorithm with a batch size of 256.

In order to cope with class imbalances, limiting the detection rate of DGAs for which only a small number of training samples exist, we apply class weighting as proposed by Tran et al.\ \cite{tran_lstm_2018}. Here, we introduce a class weight $C_{i}$ for every class $i$ which influences the magnitude of the weight updates during the training phase by weighting the loss function. Falsely classified samples of class $i$ are now penalized with $C_{i}$ instead of $1$. Therefore, the higher $C_{i}$ the more the model emphasizes on class $i$. The class weights $C_{i}$ are defined as follows:
\begin{displaymath}
C_{i} = \Big ( \frac{total\ number\ of\ samples}{number\ of\ samples\ in\ class\ i} \Big )^{\gamma}
\end{displaymath}
The parameter $\gamma$ denotes how much the data set should be rebalanced. Setting $\gamma=1$ forces the model to treat every class equally regardless of the number of samples per class included in the training data. If $\gamma=0$ is chosen, no rebalancing is applied. Similar to Tran et al.\ we propose to set $\gamma=0.3$.
\section{Evaluation Setup \& Overview}
\label{sec:evaluation}

In this section, we specify the unified setup we use in all of our comparative evaluations including the selected baseline classifiers, the data sources we used, as well as the tools and metrics applied. In addition, we motivate the experiments we conducted to address the identified open questions when bringing DGA classifiers into practice. 

\subsection{Selected State-of-the-Art Classifiers}
We first recap the currently best feature-based approach FANCI~\cite{schuppen_fanci_2018} and then continue with one candidate each for each of the two neural network types (RNN and CNN).

\subsubsection{Feature-based Approach: FANCI}
FANCI \cite{schuppen_fanci_2018} implements an RF and an SVM classifier to separate benign from malicious NXDs. In this paper, we focus on the RF classifier since it slightly outperforms the SVM approach. FANCI's classification relies on 21 hand-crafted features which can be divided into three different categories: structural, linguistic, and statistical features. The publicly accessible implementation of FANCI does not support multiclass classification off-the-shelf. While the multiclass task can easily be transformed to multiple binary classification tasks (e.g., by using multiple one-vs.-one or one-vs.-all binary classifiers), we chose not to implement multiclass classification support for FANCI. The rational is that the features used in FANCI are specifically crafted to distinguish between benign and malicious NXDs and not to distinguish between NXDs generated by different DGAs. Thus, extending FANCI to a promising multiclass classifier would require the crafting of new features which is a time-consuming task.

\subsubsection{Featureless Approaches}
All featureless approaches use the same input pre-processing as described in Section \ref{subsection:pre_processing}, and the input embedding described in Section \ref{subsection:b-resnet}. In the following, we denote classifiers which separate benign from malicious NXDs with a leading \textit{B} and classifiers which solve the multiclass classification task with a leading \textit{M}. Additionally, we differentiate between cost-sensitive and cost-insensitive models. In our experiments for the binary task, we do not evaluate cost-sensitive models since the sample distribution of the used data sets is balanced and thus the performance would not deviate from the models without class weighting. For all multiclass experiments involving cost-sensitive models (ending with \textit{.MI}), we set $\gamma=0.3$. 

\paragraph{\bf B-Endgame, M-Endgame \& M-Endgame.MI}
Woodbridge et al.\ \cite{woodbridge_predicting_2016} propose a binary and a multiclass DGA classifier which process the embedded input data with a long short-term memory (LSTM) layer containing 128 hidden units with hyperbolic tangent activation. For our evaluation, we optimized the proposed models by interchanging the standard LSTM implementation of Keras \cite{chollet_keras_2015} with the faster CuDNN~\cite{chetlur_cudnn_2014} implementation, that runs on graphics processing units (GPUs). We denote the resulting classifiers by B-Endgame and M-Endgame. Thereby, we can guarantee a fair comparison of training and classification speeds because the other deep learning models also take advantage of GPU processing. We performed several experiments which ensured that this modification has no impact on the classification performance. Further, we denote the cost-sensitive model proposed by Tran et al.\ \cite{tran_lstm_2018} by M-Endgame.MI.

\paragraph{\bf B-NYU, M-NYU \& M-NYU.MI}
Yu et al.\ \cite{yu_character_2018} propose a model based on two 1-dimensional CNN layers with 128 filters for the DGA binary classification task. We adapted the binary classifier B-NYU to a multiclass classifier M-NYU and a cost-sensitive variant M-NYU.MI using a similar approach as in M-ResNet (Section \ref{subsection:m-resnet}).

\subsection{Data Sets}
\label{sec:evaluation_data}
We use the domain names of NX-traffic from two distinct networks as sources for benign data and one source for malicious domain names generated by DGAs.
\paragraph{\bf DGArchive} Our source for malicious domains is DGArchive \cite{plohmann_comprehensive_2016}, which uses reimplementations of reverse engineered DGAs and known seeds to produce AGDs. DGArchive contains approximately 93 million unique AGDs which are produced by 88 different DGA families at the time of writing.

\paragraph{\bf University Network} We obtain benign data from the central DNS resolver of the campus network of RWTH Aachen University, which includes several academic and administrative networks, networks from student residences, eduroam~\cite{eduroam}, and the network of the university hospital of RWTH Aachen. From this source, we obtained two one-month recordings of benign NXDs. The first recording is composed of NXDs from mid-May 2017 until mid-June 2017. The second recording includes NXDs from mid-October 2017 until mid-November 2017. Each of the recordings contains around 35 million unique NXD responses.

\paragraph{\bf Company Network} The second source for benign data are several central DNS resolvers of Siemens AG\footnote{https://www.siemens.com}, which cover the regions of Asia, Europe, and the USA. From the company, we in total obtained three one-month recordings of benign NXDs. More precisely, the recordings are from Oct.~2017, Feb.~2019, and Apr.~2019, and contain approximately 20 million, 22 million, and 27 million unique NXDs, respectively.
\\\\
These data sources provide us with diverse data of different times of days and different days of a week including working and non-working days. Therefore, these recordings allow for the creation of representative real-world data sets. Primarily, the benign non-resolving domain names originate from typing errors caused by humans, faulty or outdated software trying to resolve domains that do not exist, or by the intentional misuse of the DNS for different purposes. To further clean our benign sets, we filter all university and company recordings against DGArchive and remove all known malicious AGDs from the benign labeled data. However, benign AGDs, such as domains generated by the DGA included in Google Chrome used for DNS hijacking detection \cite{zdrnja_google_2011}, or domains generated by antivirus software for signature checks \cite{sophos_sophos_2019}, remain and are accounted as benign.

\subsection{Experimental Overview \& Metrics}
\label{sec:evaluation_expermients}
All deep learning based classifiers are evaluated on an NVIDIA Tesla V100 GPU using Python 3.6.0, Keras 2.24, TensorFlow 1.13.1, CUDA 10.0.130, and cuDNN 7.4.

We divide our evaluation of the classifiers into two parts. In the first part (Section~\ref{sec:classification_performance}), we evaluate the classification performance of the classifiers. In the second part (Section~\ref{sec:properties}), we compare the classifiers with respect to the explainability of their predictions, their robustness against two adversarial attacks and their training and classification speed. In addition, we evaluate B-ResNet's generalizability with respect to changes in time and network environment.

The evaluation of the classification performance of the classifiers is further divided into the following three parts:

\begin{table}[!t]
	\caption{Evaluation Metrics}
	\label{tab:metrics}
	\centering
	\begin{tabular}{ll}
		\toprule
		$ ACC = \frac{TP + TN}{TP + TN + FP + FN} $ & $Precision = \frac{TP}{TP + FP}$ \\
		\addlinespace[0.16cm]
		$Recall = \frac{TP}{TP + FN}$ & $F1{-}score = 2 \cdot \frac{Precision \cdot Recall}{Precision + Recall}$ \\
		\bottomrule
	\end{tabular}
\end{table}

In Section \ref{sec:binary_classification}, we assess the performance of the binary classifiers. We start with  what we refer to as the \emph{mixed DGAs} setting in which we assess the classifiers' ability to correctly classify AGDs of known DGAs. I.e., we train the classifiers on samples of known DGAs and evaluate the performance on different samples generated by the same DGAs. 

In a second binary classification experiment, we test the capability of the classifiers to detect AGDs of unknown DGAs. Here, we evaluate the detection rate on AGDs which are generated by a DGA for which no samples were included in the training~set. 

In both experiments, the accuracy (ACC) is used as primary metric to measure the performance of the classifiers. The ACC is computed from the total number of true positives $TP$, true negatives $TN$, false negatives $FN$, and false positives $FP$ as shown in Table~\ref{tab:metrics}. As additional metrics, we provide the true positive rate (TPR), true negative rate (TNR), false negative rate (FNR), and the false positive rate (FPR). 

In Section \ref{sec:multiclass_classification}, the performance of the multiclass classifiers are compared. Here, we assess the performance of the classifiers by the f1-score, recall, and precision. The formulas for the calculation of these metrics are included in Table~\ref{tab:metrics}. For computing the overall evaluation metrics we chose macro-averaging over micro-averaging as we want to analyze the effect of class imbalance on different classifiers. The former averages the evaluation metrics over all classes regardless of the number of samples per class inside the test set, giving each class the same level of importance. The latter accounts each sample an equal level of importance making the overall score biased towards well-represented classes.
\section{Classification Performance}
\label{sec:classification_performance}

This section is dedicated to the comparative evaluation of the classification performance of the various binary and multiclass classifiers, including the detection of unknown DGAs.
 
\subsection{Binary Classification}
\label{sec:binary_classification}

Since we strive for a fair comparison of deep learning approaches with a classical feature-based classifier, we first reproduce the original paper results of FANCI and in the second step we evaluate the deep learning classifiers on the same data. For the binary evaluation, we use 20 data sets with a set size of 92,102 samples which contain as many benign samples as malicious samples. The malicious samples are uniformly distributed among all DGAs included in the data sets of \cite{schuppen_fanci_2018}. All samples are drawn uniformly at random from the one-month May/June recording of the university benign source and from the DGArchive data for malicious samples.

\subsubsection{Mixed DGAs}
\label{sec:binary_classification_mixed_dga}
We investigate the capability of the classifiers to detect AGDs of known DGAs by performing 5 repetitions of a 5-fold cross validation on the 20 previously described data sets. Accordingly, for each fold we split the data sets into 80\% training and 20\% testing samples. Additionally, we remove 5\% of the training data for a holdout set, which is used to validate the performance of the deep learning classifiers during training. We train the deep learning models until there are no further improvement on the holdout set.

\begin{table}[!t]
	\caption{Binary Classification: Mixed DGAs}
	\label{tab:mixed_dga}
	\centering
	\small
	\begin{tabular}{lccccl}
		\toprule
		\textbf{Classifier} & \textbf{ACC} & \textbf{TPR} & \textbf{TNR} & \textbf{FNR} & \textbf{FPR} \\
		\midrule
		FANCI & 0.99764 & 0.99744 & 0.99784 & 0.00256 & 0.00216 \\
		B-Endgame & 0.99891 & 0.99969 & 0.99813 & 0.00031 & 0.00187 \\
		B-NYU & 0.99907 & 0.99976 & 0.99838 & 0.00024 & 0.00162 \\
		B-ResNet & 0.99916 & 0.99978 & 0.99853 & 0.00022 & 0.00147 \\
		\bottomrule
	\end{tabular}
\end{table}

The averaged results for the different classifiers are depicted in Table \ref{tab:mixed_dga}. Each classifier offers an excellent detection rate. FANCI achieves a similar performance as stated in \cite{schuppen_fanci_2018}. However, all deep learning based classifiers perform slightly better than FANCI. The FNR even decreases by one order of magnitude compared to FANCI. Regarding the achieved accuracy, our B-ResNet classifier performs at least comparable to if not better than the other deep learning classifiers and shows the smallest FPR.

\subsubsection{Unknown DGAs}
\label{sec:binary_classification_unknown_dga}
Here, we evaluate the capability of the classifiers to detect AGDs of unknown DGAs. To this end, we perform 5 repetitions of a leave-one-group-out cross validation for each of the 20 data sets. We train the classifiers on samples of all DGAs but one, and evaluate the performance on the AGDs of the left out~DGA.

\begin{table}[!t]
	\caption{Binary Classification: Unknown DGAs}
	\label{tab:logo_dga}
	\centering
	\small
	\begin{tabular}{lccccl}
		\toprule
		\textbf{Classifier} & \textbf{ACC} & \textbf{TPR} & \textbf{TNR} & \textbf{FNR} & \textbf{FPR} \\ 
		\midrule
		FANCI & 0.98073 & 0.96392 & 0.99754 & 0.03608 & 0.00246  \\
		B-Endgame & 0.98569 & 0.97307 & 0.99830 & 0.02693 & 0.00170 \\
		B-NYU & 0.98406 & 0.96961 & 0.99851 & 0.03039 & 0.00149 \\
		B-ResNet & 0.98517 & 0.97168 & 0.99866 & 0.02832 & 0.00134 \\
		\bottomrule
	\end{tabular}
\end{table}
The averaged results are depicted in Table \ref{tab:logo_dga}. All classifiers are highly capable of detecting unknown DGAs. All deep learning approaches perform slightly better than FANCI but among themselves they perform approximately equally well.

\subsection{Multiclass Classification}
\label{sec:multiclass_classification}
In this section, we compare the performance of the classifiers according to the multiclass classification task. Here, we measure the ability of the classifiers to correctly label a sample either as benign, or in the case of an AGD, with the corresponding DGA family which generated the domain name. 

To this end, we create a data set which contains 10,000 random samples for each DGA in DGArchive for which at least this many samples exist and all samples of DGAs for which less than 10,000 samples are known. Additionally, we include 10,000 random samples of the Apr. 2019 recording of the company benign data source. This results in 89 classes including the benign class and yields an overall set size of 526,534 samples. 

For evaluation, we perform 5 repetitions of a 5-fold cross validation where we split the samples of each included class into 80\% training and 20\% testing samples in every fold.

\begin{table}[!t]
	\caption{Multiclass Classification}
	\label{tab:mcc_unbalanced}
	\centering
	\small
	\begin{tabular}{lccl}
		\toprule
		\textbf{Classifier} & \textbf{F1-score} & \textbf{Precision} & \textbf{Recall} \\
		\midrule
		M-Endgame & 0.72541 & 0.74319 & 0.72567 \\
		M-Endgame.MI & 0.74312 & 0.76022 & 0.74499 \\
		M-NYU & 0.68913 & 0.71867 & 0.68738 \\
		M-NYU.MI & 0.73832 & 0.76104 & 0.73993 \\
		M-ResNet & 0.78878 & 0.81734 & 0.78850 \\
		M-ResNet.MI & 0.79648 & 0.81266 & 0.80306 \\
		\bottomrule
	\end{tabular}
\end{table}

\begin{figure}[!t]
	\centering
	\includegraphics[width=\linewidth]{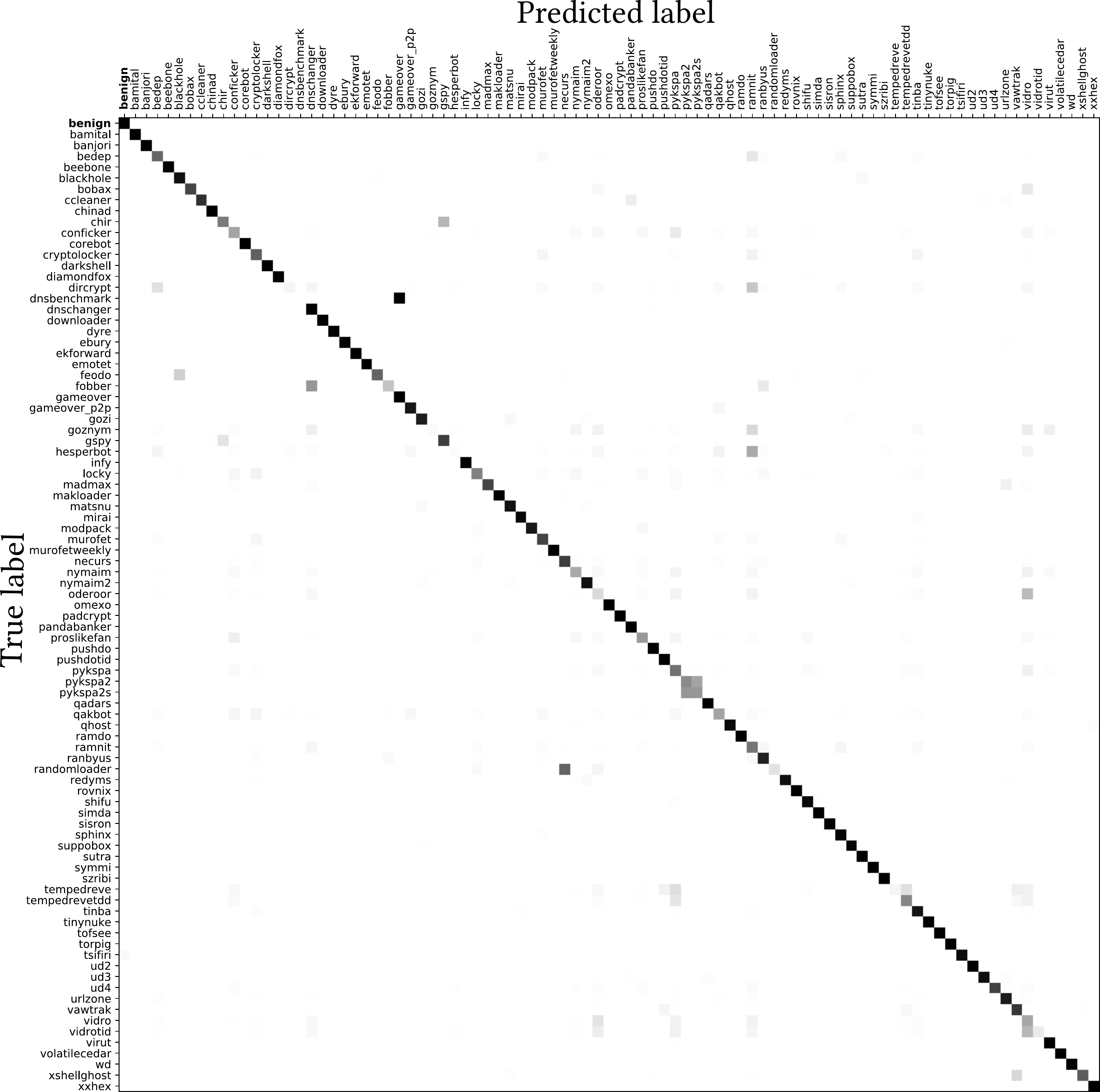}
	\caption{Confusion matrix of M-ResNet.MI.}
	\label{fig:unbalanced}
\end{figure}

Table \ref{tab:mcc_unbalanced} shows the averaged results for the different classifiers. The cost-sensitive approaches always perform better than their cost-insensitive counterpart. The M-NYU model profits most while the improvements of M-ResNet are minor. M-ResNet.MI outperforms every other model and achieves an f1-score which is more than 5 percentage points above the next best state-of-the-art classifier. Recall that we use macro-averaging for computing the overall f1-score. An improvement of 5\% means that our classifier is able to correctly classify AGDs of several classes which the other classifiers are not able to correctly attribute. For example, for the 7 classes for which we observe the largest difference in f1-scores between M-ResNet.MI and M-Endgame.MI, the former achieves an averaged f1-score of 0.80322 and the latter one of 0.19917. On average, for each of theses DGAs only 40 samples are available for training, indicating that ResNets perform particularly well in the extraction of features from smaller amounts of samples.

To visualize the classification performance of M-ResNet.MI we additionally provide the corresponding confusion matrix in Fig.~\ref{fig:unbalanced}. The blocks inside the figure depict the fraction of samples of the classes on the vertical axis which are labeled as classes on the horizontal axis. 100\% is represented by a clear black block and 0\% is represented as a clear white block. A perfect classifier would yield an identity matrix of black blocks. It can be seen that the benign class (located in the top-left corner) can be precisely separated from all other classes (f1-score of 0.99684). Most DGA families are easily recognizable. The most striking outlier is \textit{Dnsbenchmark}. 100\% of its test samples are labeled as \textit{Gameover}. This is due to the small amount of available AGDs for this DGA. The training set includes only 4 samples and the classification score for the whole class relies on the classification of one single sample. Hence, when this single sample is labeled incorrectly, the f1-score of the whole class degrades to zero in the corresponding fold. The same applies to the \textit{Randomloader} class, which also contains only 5 samples. However, here, the classifier is able to label some of its AGDs correctly. In general, DGA families which are related to each other such as \textit{Pykspa2} and \textit{Pykspa2s} generate similar AGDs. Therefore, it is hard for the classifier to separate these classes among themselves but it is still able to delimit them from all other classes.

\section{Classifier Properties}
\label{sec:properties}
We compare the classifiers according to the explainability of their classification results, their robustness against two evasion strategies, and their training and classification speed. We also evaluate how well B-ResNet generalizes to different network environments and with respect to time.

\subsection{Explainability}
While the predictions of feature-based approaches such as FANCI can be traced down to the individual features of AGDs, this is not possible for the deep learning based classifiers. Therefore, in this section, we strive to explain the predictions of these approaches by investigating the falsely attributed AGDs from the multiclass classification experiment (Section \ref{sec:multiclass_classification}). To highlight some of the self-learned features used by the deep learning classifiers, we extract simple regular expressions for each DGA and check it against all samples of this DGA included in DGArchive.

\begin{table}[!t]
	\caption{Regular Expressions for Several DGAs Grouped by Similarity}
	\label{tab:agds_regex}
	\centering
		\resizebox{\columnwidth}{!}{
		\begin{tabular}{lrr}
			\toprule
			\textbf{DGA} & \textbf{\#Training Samples} & \textbf{AGD Regex} \\
			\midrule
			bedep & 5966 & [a-z0-9]\{12,18\}.com \\
			dircrypt & 552 & [a-z]\{8,20\}.com \\
			dnschanger & 8000 & [a-z]\{10\}.com \\
			goznym & 291 & [a-z]\{5,12\}.com \\
			hesperbot & 142 & [a-z]\{8,24\}.com \\
			ramnit & 8000 &[a-y]\{8,19\}.(bid$\vert$click$\vert$com$\vert$eu) \\
			\midrule
			feodo & 153 & ([a-z]\{16\}$\vert$[a-z]\{18\}).ru \\
			blackhole & 585 & [a-z]\{16\}.ru \\
			\midrule
			oderoor & 8000 & [a-z]\{7,12\}.(cc$\vert$com$\vert$dyndns.org$\vert$net$\vert$tv) \\
			vidro & 8000 & [a-z]\{7,12\}.(com$\vert$dyndns.org$\vert$net) \\
			\bottomrule
	\end{tabular}
	}
\end{table}

Table \ref{tab:agds_regex} shows an excerpt of these expressions grouped by similarity including the number of training samples per DGA. For instance, the regular expression for \textit{Ramnit} is \mbox{[a-y]\{8,19\}.(bid$\vert$click$\vert$com$\vert$eu)}. I.e., this DGA generates strings of length 8 to 19 using only characters from \mbox{a to y} and then appends the dot and one of the four possible TLDs to the generated strings yielding the final AGDs.\footnote{87.8\% of all \textit{Ramit's} AGDs in DGArchive end with the .com TLD}

As illustrated in the confusion matrix of M-ResNet.MI (Fig.\ref{fig:unbalanced}), a huge fraction of the AGDs belonging to \textit{Bedep}, \textit{Dircrypt}, \textit{Goznym}, and \textit{Hesperbot} are falsely labeled as \textit{Ramnit}. The latter three DGAs are generally very poorly detected by M-ResNet.MI yielding nearly clear white diagonal cells in the confusion matrix. Considering that a huge fraction of the AGDs generated by these three DGAs match the same regular expression, this is not surprising. Although, class weighting with $\gamma=0.3$ is utilized the classification is still biased towards the better represented classes, in this case towards \textit{Ramnit}.

By investigating which specific samples of this group of DGAs are falsely labeled we can draw conclusions on the possible features used for classification. For instance, within the samples of \textit{Bedep} which are misclassified by M-ResNet.MI as \textit{Ramnit}, there are no samples which contain numbers or the character ``z''. Consequently, we reckon that the set of characters included within AGDs is used as features to separate different DGA families in M-ResNet.MI. In comparison, M-Endgame.MI does also not classify any samples of \textit{Bedep} which include numbers as \textit{Ramnit} but incorrectly labels samples which include the character ``z''. The corresponding set of falsely labeled samples of M-NYU.MI contains both, samples with numbers and samples with ``z''. 

We also reckon that the AGD length is used as a feature for classification. Investigating the samples of \textit{Goznym} which are classified as \textit{Ramnit} reveals that M-ResNet.MI and M-Endgame.MI do not wrongly classify any samples shorter than 8 characters or containing the character ``z''. Similarly, M-NYU.MI does not label \textit{Goznym's} samples shorter than 8 characters as \textit{Ramnit} but fails to separate AGDs of those classes based on the character ``z''. The separation based solely on the length of AGDs is observable on the samples generated by \textit{Feodo} which are classified as \textit{Blackhole}, since no samples of length 18 are mislabeled.

Lastly, we reckon that the TLD of AGDs is used to separate different DGA families. Investigating the samples generated by \textit{Oderoor} which are labeled as \textit{Vidro} reveals that no AGDs which use the .cc or .tv TLD are misclassified.

\subsubsection{Complex Feature Extraction Capability}
So far we only identified features potentially learned by the deep learning classifiers to discriminate between different DGA families which are also easily recognizable by a human. In the following, we examine the capability of the classifiers to extract more complex features. To this end, we use implementations of reverse engineered DGAs from \cite{bader_dgas} to generate a data set which incorporates samples of two DGAs which are not distinguishable via their AGD regular expression. Specifically, we use the \textit{DNSchanger} DGA which generates domain names that match \mbox{[a-z]\{10\}.com}, and adjusted the DGA \textit{Dircrypt} such that the AGDs it generates match the same regular expression by fixing the length of the strings it generates to 10 (see Table \ref{tab:agds_regex}). Thereby, we create a data set composed of 350,000 unique samples per class. We then perform 5 repetitions of a 5-fold cross validation to assess the performance of the different binary classification models.

\begin{table}[!t]
	\caption{Complex Feature Extraction Capability}
	\label{tab:PRNG-eval}
	\centering
	\small
	\begin{tabular}{cccc}
		\toprule
		\multicolumn{4}{c}{\textbf{Accuracy}} \\
		\textbf{FANCI} & \textbf{B-Endgame} & \textbf{B-NYU} & \textbf{B-ResNet} \\
		\midrule
		0.50188 & 0.58441 & 0.51569 & 0.69904\\
		\bottomrule
	\end{tabular}
\end{table}

Table~\ref{tab:PRNG-eval} shows the averaged results for the different classifiers. It is not surprising that FANCI fails to separate the two classes (ACC of 50.188\%) considering that its features are engineered to discriminate between benign and malicious NXDs and not to distinguish different DGAs. However, the performance of B-NYU (ACC of 51.569\%) is also close to random guessing. B-Endgame achieves an accuracy of 58.441\%, but B-ResNet outperforms all other approaches with an accuracy of 69.904\%. B-ResNet is thus able to label a huge fraction of samples based on the used generation algorithm with a probability significantly higher than random guessing. 

This experiment indicates that B-ResNet is better suited to extract complex features than any of the previously proposed classifiers. We reckon that this is due to its architecture and choice of hyperparameters.

\subsection{Robustness}
We assess the robustness of the different classifiers by measuring their detection capabilities when presented with samples generated by one of two evasion strategies. These have recently be shown to be effective against FANCI and B-Endgame in the case where these classifiers were trained using Alexa domains or resolving DNS traffic as benign data.

Peck et al.\ propose \textit{Charbot}\cite{peck_charbot_2019}, a DGA which randomly selects a second level domain from the Alexa top domain names, replaces two random characters and appends one of 22 TLDs in order to generate an AGD. Spooren et al.\ iteratively developed the \textit{Deception} DGA\cite{spooren_detection_2019} targeting FANCI by exploiting the handcrafted features used for classification. Thereby, the authors crafted a DGA which specifically aims to fool FANCI but is at the same time difficult to detect by deep learning based approaches.

For this evaluation, we obtain 150,000 samples of the \textit{Deception} DGA from DGArchive and generate the same amount of samples using the \textit{Charbot} algorithm. Since the \textit{Deception} samples do not contain any TLDs, we randomly append one of the TLDs used by \textit{Charbot} to each sample. For both DGAs, we create one individual data set by including 150,000 benign samples drawn uniformly at random from the Oct./Nov. 2017 university data. We then use the trained binary classifiers from the \textit{mixed DGA} experiment (Section~\ref{sec:binary_classification_mixed_dga}) to evaluate the classifiers' performances against the adversarial examples. 

\begin{table}[!t]
	\caption{Robustness Analysis}
	\label{tab:robustness_analysis}
	\centering
	\resizebox{\columnwidth}{!}{
		\begin{tabular}{lcccc}
			\toprule
			\multicolumn{5}{c}{\textbf{Benign data: NX-traffic}} \\
			\textbf{Classifier} & \multicolumn{2}{c}{\textbf{TPR at FPR=0.01}} & \multicolumn{2}{c}{\textbf{TPR at FPR=0.001}} \\
			\cmidrule(lr){2-3}
			\cmidrule(lr){4-5}
			& \textbf{charbot} & \textbf{deception} & \textbf{charbot} & \textbf{deception} \\
			\midrule
			FANCI & 0.98064 & 0.99994 & 0.94283 & 0.98919\\
			B-Endgame & 0.98644 & 0.99952 & 0.80812 & 0.80575\\
			B-NYU & 0.99017 & 0.99987 & 0.94053 & 0.95185\\
			B-Resnet & 0.98793 & 0.99958 & 0.92144 & 0.93696\\
			\midrule
			\multicolumn{5}{c}{\textbf{Peck et al.\cite{peck_charbot_2019}, benign data: resolving DNS traffic / Alexa data}} \\
			\midrule
			FANCI & 0.2143 / 0.0305 & 0.4685 / 0.0166 & - / - & - / - \\
			B-Endgame & 0.3190 / 0.1550 & 0.3773 / 0.1274 & 0.1525 / 0.0558 & 0.1661 / 0.0402 \\
			\bottomrule
		\end{tabular} 
	}
\end{table}

Table \ref{tab:robustness_analysis} visualizes the averaged TPRs of 20 passes per classifier and DGA at an FPR of 1\% and 0.1\%. Recall that in our setup all classifiers are trained on real-world NXDs as benign samples. In comparison, Table \ref{tab:robustness_analysis} also includes the TPRs of FANCI and B-Endgame reported in \cite{peck_charbot_2019}, using resolving DNS traffic or Alexa top domain names as source for benign samples. Our evaluation shows that all classifiers are remarkably robust against both adversarial attacks when NX-traffic is used as benign data for training. At a fixed FPR of 1\% the detection rates for \textit{Charbot} (98\%-99\%) are slightly lower compared to the TPRs for the \textit{Deception} DGA, where nearly all samples are labeled correctly. At a fixed FPR of 0.1\% the detection rates of B-Endgame are by far the worst. Interestingly, FANCI achieves the highest TPRs for the \textit{Deception} DGA although the DGA was specifically created to evade detection by FANCI.

In comparison, the TPRs of B-Endgame and FANCI trained on Alexa top domains or resolving traffic reported in \cite{peck_charbot_2019} are significantly lower. The authors of \textit{Charbot} were not even able to establish a classification threshold for FANCI achieving an FPR of 0.1\%, probably due to FANCI's features which focus on separating benign and malicious NXDs. As could be expected, classifiers which are trained on Alexa top domain names are easier to fool. \textit{Charbot} uses the Alexa list as basis for its AGD generation, whereas during the iterative development of the \textit{Deception} DGA the Alexa top domain names served as benign samples for the targeted classifier. Similarly, the huge differences between the performances of the classifiers when training is performed on NX-traffic can be explained by the fact that the domain names of the resolving DNS traffic, which are used for training, are naturally more similar to the Alexa top domain names that are used to create the adversarial examples. 

While training on NXDs does not enhance robustness against adversarial attacks in general, it increases the effort required to generate adversarial AGDs as an attacker has to be in possession of appropriate NX-traffic from a sufficiently large network in order to craft such a DGA.

\subsection{Generalization}
We evaluate how well B-ResNet generalizes to different networks and thus in how far the classification can be outsourced as a service. To this end, we train B-ResNet on data recorded in one network and perform classification on samples observed in the other network. Subsequently, we investigate whether frequent retraining is necessary in order to adjust to possible changes within a network over time. Here, we train the classifier on data which is available up to a certain point in time and then evaluate its performance upon prediction of samples which were recorded at a later point in time.

\subsubsection{Network Generalization}
We show that the classification performance of B-ResNet is independent of the specific network it was trained in. To this end, we perform experiments similar to the ones described in Section \ref{sec:binary_classification_mixed_dga} in which we detect arbitrary AGDs of known DGAs. For the sake of clarity in the tables, we omitted statistical values except for the averages $\overline{x}$ when comparing the classifiers. In order to be able to analyze the generalizability of B-ResNet in detail, we here additionally provide the standard deviation $\sigma$, the minimum $x_{min}$ and maximum $x_{max}$ values of the scores, and the median $\tilde{x}$ for B-ResNet in the upper part of Table~\ref{tab:generalization_combined}.

\begin{table}[!t]
	\caption{Generalization Experiments}
	\label{tab:generalization_combined}
	\centering
	\resizebox{\columnwidth}{!}{
		\begin{tabular}{llccccl}
			\toprule
			\textbf{Scenario} & \textbf{} & \textbf{ACC} & \textbf{TPR} & \textbf{TNR} & \textbf{FNR} & \textbf{FPR} \\
			\midrule
			\multirow{2}{*}{Baseline:} & $\overline{x}$ & 0.99916 &	0.99978 & 0.99853 &	0.00022 & 0.00147 \\
			\multirow{2}{*}{B-ResNet,} & $\sigma$  & 0.00013 &	0.00014 & 0.00024 &	0.00014 & 0.00024 \\
			\multirow{2}{*}{Mixed DGAs} & $x_{min}$ & 0.99852 &	0.99928 & 0.99739 &	0.00000 & 0.00102 \\
			\multirow{2}{*}{(Section \ref{sec:binary_classification_mixed_dga})}& $\tilde{x}$ & 0.99917 &	0.99983 & 0.99855 &	0.00017 & 0.00145 \\
			& $x_{max}$ & 0.99938 &	1.00000 & 0.99898 &	0.00072 & 0.00261 \\
			\midrule
			\midrule
			\multirow{2}{*}{Network}& $\overline{x}$ & 0.99761 & 0.99991 & 0.99531 & 0.00009 & 0.00469  \\
			\multirow{2}{*}{Generalization:}& $\sigma$ & 0.00034 & 0.00005 & 0.00070 & 0.00005 & 0.00070  \\
			\multirow{2}{*}{Train University,}& $x_{min}$ & 0.99693 & 0.99982 & 0.99386 & 0.00000 & 0.00354  \\
			\multirow{2}{*}{Test Company}& $\tilde{x}$ & 0.99770 & 0.99991 & 0.99544 & 0.00009 & 0.00456  \\
			& $x_{max}$ & 0.99817 & 1.00000 & 0.99646 & 0.00018 & 0.00614  \\
			\midrule
			\multirow{2}{*}{Network} & $\overline{x}$ & 0.99869 & 0.99979 & 0.99759 & 0.00021 & 0.00241 \\
			\multirow{2}{*}{Generalization:} & $\sigma$ & 0.00023 & 0.00011 & 0.00051 & 0.00011 & 0.00051 \\
			\multirow{2}{*}{Train Company,} & $x_{min}$ & 0.99834 & 0.99955 & 0.99678 & 0.00004 & 0.00155 \\
			\multirow{2}{*}{Test University} & $\tilde{x}$ & 0.99865 & 0.99982 & 0.99761 & 0.00018 & 0.00239 \\
			& $x_{max}$ & 0.99903& 0.99996 & 0.99845 & 0.00045 & 0.00322 \\
			\midrule
			\midrule
			\multirow{2}{*}{Time} & $\overline{x}$ & 0.99887 & 0.99982 & 0.99793 & 0.00018 & 0.00207 \\
			\multirow{2}{*}{Generalization,} & $\sigma$ & 0.00009 & 0.00014 & 0.00022 & 0.00014 & 0.00022 \\
			\multirow{2}{*}{Difference:}& $x_{min}$ & 0.99867 & 0.99942 & 0.99737 & 0.00003 & 0.00175 \\
			\multirow{2}{*}{1 Month} & $\tilde{x}$ & 0.99887 & 0.99985 & 0.99798 & 0.00015 & 0.00202 \\
			& $x_{max}$ & 0.99902 & 0.99997 & 0.99825 & 0.00058 & 0.00263 \\
			\midrule
			\multirow{2}{*}{Time} & $\overline{x}$ & 0.99777 & 0.99774 & 0.99779 & 0.00226 & 0.00221 \\
			\multirow{2}{*}{Generalization,} & $\sigma$ & 0.00062 & 0.00106 & 0.00089 & 0.00106 & 0.00089 \\
			\multirow{2}{*}{Difference:} & $x_{min}$ & 0.99634 & 0.99506 & 0.99477 & 0.00076 & 0.00172 \\
			\multirow{2}{*}{17 Months} & $\tilde{x}$ & 0.99784 & 0.99781 & 0.99807 & 0.00219 & 0.00193 \\
			& $x_{max}$ & 0.99863 & 0.99924 & 0.99828 & 0.00494 & 0.00523 \\
			\bottomrule
		\end{tabular}
	}
\end{table}

We create 20 data sets for each of the two benign data sources, the university and the company network. Each of these 40 sets includes 1,000 randomly picked samples of each DGA in DGArchive for which that many AGDs are known and as many samples as possible if less than 1,000 AGDs are available. As benign data, each of the sets contains as many randomly chosen samples from the respective benign data source as malicious samples are included. The benign samples were recorded approximately at the same point in time in both networks, more precisely Oct./Nov. 2017. For each of the 40 sets, we train a classifier and perform classification on all 20 sets which include benign samples of the respective other network. Thus, we perform $20 \cdot 20 = 400$ evaluations for training on university data and testing on company data and $400$ evaluations for the other direction.  

The middle part of Table \ref{tab:generalization_combined} shows the averaged results for the network generalization experiment. When training on university data and predicting on company data, the ACC slightly decreases and the standard deviation increases for every metric except for the TPR and the FNR  compared to the baseline results. The increase of the FPR is an expected outcome since benign samples of one network may miss specific properties of samples from the other network. The best ACC is achieved at an FNR of 0.00009 which is less than half of the baseline's FNR. Therefore, it might be possible to decrease the FPR by moving the decision threshold at the expanse of the FNR and ACC. 

For the other direction (training on company data and testing on university data), the ACC again decreases slightly compared to the baseline. However, this time the FNR stays in the same order of magnitude and the increase of the FPR is smaller. 

Both of the experiments show that B-ResNet is able to generalize very well to unknown networks.

\subsubsection{Time Generalization}
\label{sec:time_generalization}
We analyze whether it is advisable to frequently retrain B-ResNet. We perform two evaluations in which we train the classifiers using only DGArchive and company data that was recorded up to a specific point in time. Hence, only AGDs of DGAs which were present in DGArchive at that time are included in the training sets. First, we train the classifier on data which was available up to Mar. 2019 and evaluate on data which was recorded in Apr. 2019. Thereby, we have at least a one-month time difference between the training and testing samples. Then, we repeat this experiment with a time difference of 17 months (train on data available up to Nov. 2019 and evaluate on Apr. 2019 data). Similarly to the network generalization experiments, we create 20 sets per time interval yielding 400 evaluation passes per time difference. 

The lower part of Table \ref{tab:generalization_combined} depicts the results of the time generalization experiment. The results for a time distance of one month are very similar to the ones of the baseline. The FPR is slightly higher which could be due to changes in the benign data caused by changes within the company's network. The FNR stays in the same order of magnitude. Since the ACC is nearly as high as for the baseline and the standard deviation is small, it seems unnecessary to retrain the classifier after one month.

The results for the prediction on data using a classifier which was trained on samples recorded 17 months ago show an increased standard deviation of the FPR compared to the previous experiment. Again, this might be caused by changes within the network. Moreover, the FNR increases by more than a factor of 10 compared to the baseline. The reason for this is that the 4 DGAs: \textit{Ccleaner}, \textit{Nymaim2}, \textit{Tinynuke}, and \textit{Wd} are not included in the training sets but in the testing sets. The averaged TPR for these DGAs is 0.96790 while the TPR for \textit{Wd} is the worst with 0.89690. However, the AGDs generated by \textit{Ccleaner} and \textit{Tinynuke} can be detected with a TPR of 1.00000 and 0.99998, respectively. Therefore, while retraining is not required for a high detection rate for some DGAs such as \textit{Ccleaner} and \textit{Tinynuke}, the TPR for other DGAs such as \textit{Wd} can be significantly increased by including samples of it into the training set. Excluding these four DGAs from the TPR calculation results in an overall score of 0.99981 with a standard deviation of 0.00010. 

These results show that B-ResNet is remarkably time-robust and  able to detect AGDs of known DGAs independent of the generation date of the AGD. Thus, retraining B-ResNet may be useful when new DGAs are detected in order to decrease the FNR. To achieve the smallest possible FPR, we recommend retraining the classifier as frequently as possible.

\subsection{Training \& Classification Speed}
Here, we compare the training and classification speed of the featureless classifiers. We intentionally exclude FANCI from this comparison as all other approaches are evaluated on a GPU while FANCI is executed on the CPU. For the comparison, we measure the training and classification time in the binary mixed DGAs experiment (Section~\ref{sec:binary_classification_mixed_dga}) and in the multiclass  experiment (Section~\ref{sec:multiclass_classification}).

\begin{table}[!t]
	\caption{Performance Analysis}
	\label{tab:speed}
	\centering
	\small
	\begin{tabular}{lcc}
		\toprule
		\textbf{Classifier} & \boldmath$\frac{Training\ Time}{Classifier} [s]$ & \boldmath$\frac{Classification\ Time}{Sample} [\mu s]$ \\
		\midrule
		B-Endgame & 103.8 & 82.6  \\
		B-NYU & 33.0 & 22.5  \\
		B-ResNet & 36.5 & 32.6  \\
		\midrule
		M-Endgame & 1310.3 & 70.9 \\
		M-Endgame.MI & 1268.8 & 66.6 \\
		M-NYU & 446.1 & 17.3 \\
		M-NYU.MI & 410.4 & 17.5 \\
		M-ResNet & 871.6 & 73.1 \\
		M-ResNet.MI & 886.6 & 72.3 \\
		\bottomrule
	\end{tabular}
\end{table}

Table \ref{tab:speed} depicts the averaged time to train one classifier and the classification times per sample over all passes for both experiments and each of the deep learning classifier. For the binary classification task, the B-Endgame classifier requires the most time for training as well as classification. B-NYU and B-ResNet are similar with respect to training times but the NYU model needs 10.1 $\mu s$ less time to classify a single sample. In the multiclass scenario, the NYU classifiers are the fastest to train. The ResNet classifiers require approximately twice the training time of the NYU models and the Endgame classifiers need thrice as much time. For classification, the NYU models need by far the least time. The Endgame and the ResNet-based models classify samples at a similar speed.

Note that the ResNet-based approaches are fast enough to perform live detection in large networks. Within the network of the company, there are on average 148 NXD responses per second with a maximum peak of 2471 NXDs per second. B-ResNet classifies a single sample in $32.6\mu s$ and therefore is able to classify 30,674 samples per second. M-ResNet.MI needs $72.3\mu s$ for a single domain name and therefore classifies 13,831 samples per second. Hence, the ResNet-based approaches are real-time capable and can perform live detection in large networks for both classification tasks. 

\begin{figure}[!t]
	\centering
	\includegraphics[width=0.816\linewidth]{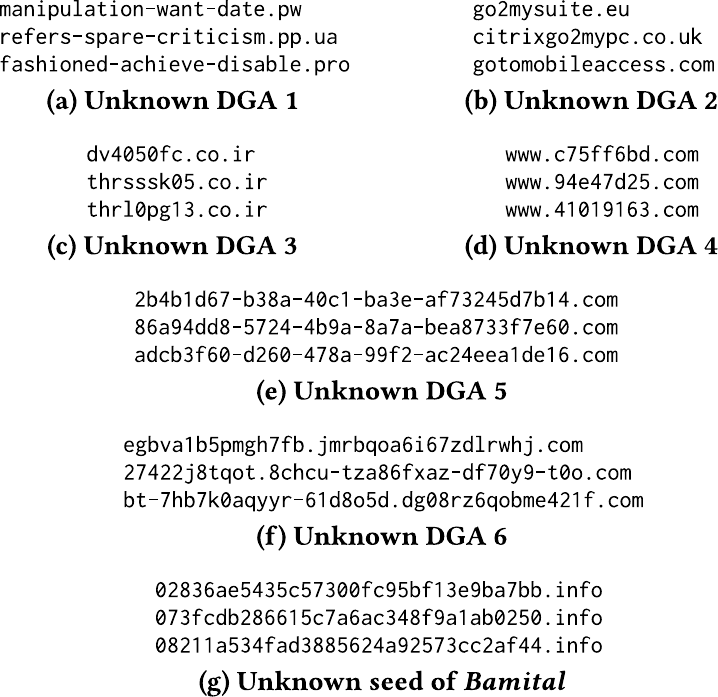}
	\caption{Real world samples of unknown AGDs.}
	\label{fig:new_dgas}
\end{figure}

\section{Real World}
To test B-ResNet's real-world applicability, we reuse the 20 classifiers of the time generalization experiment (Section~\ref{sec:time_generalization}). Recall that these are trained on a subset of the Feb. 2019 company data~and DGArchive data up to March 2019. Each classifier is used to classify the whole unfiltered recording of Apr. 2019 in the company network. The results are then averaged. In total, the recording comprises over 370 million NXD responses, 27 million of which are unique.

On average, the classifiers labeled 69,138 unique samples as malicious. Filtering the whole recording against DGArchive, we observe 13,870 unique AGDs which are generated by 6 different known DGAs. This is what a traditional blacklist would detect. Notably, 17 of the 20 classifiers are able to detect all known AGDs and the average TPR is 99.997\%. Through a semiautomatic examination of the remaining positives we try to reveal unknown AGDs. Note that ``unknown'' in this context means that the AGDs are not included in DGArchive and could not be found in other common OSINT sources, such as VirusTotal, at the time of writing. To find groups within unknown AGDs we cluster them by their length, included characters, TLDs, number of queries, time span (from the first occurrence to the last one), Shannon entropy, and number of included English words. We label the groups, either as unknown DGAs or as unknown seeds of a known DGA, with the help of DGArchive, domain knowledge, and manual research. 

With this technique we are able to report 5,833 unknown AGDs, which are represented by 8 clusters of which we reckon 6 to be unknown DGAs, 1 an unknown seed of \textit{Bamital}, and 1 to be the \textit{Conficker} DGA. A further examination of the found \textit{Conficker} AGDs revealed, that these domains are generated several months ahead of their validity period, explaining why these samples were not included in DGArchive. Possible explanations for this could be that the re-implementation which is used to create the blacklist for this DGA is faulty, the infected hosts have incorrect time settings, or this could be a new \textit{Conficker} DGA variant which intentionally generates AGDs ahead of its original validity period in order to circumvent simple blacklisting approaches. Fig.~\ref{fig:new_dgas} shows representatives of unknown AGDs for each of the remaining groups.

Concluding, within this one-month time period we reckon 49,435 unique samples to be false positives, which corresponds to an FPR of 0.00182 and approximately 69 false positives per hour for a large international company. Note that there is a one-month time difference between training the classifiers and performing the prediction. As indicated in the time generalization experiment (Section$  $ \ref{sec:time_generalization}), a more recently trained classifier would achieve an even lower FPR. In practice, the intrusion detection system which incorporates DGA detection should only fire an alert if for a host the number of positively marked samples exceeds a configurable threshold. Thereby, the false alarm rate can be reduced significantly.
This real-world application test shows the practicability~of~\mbox{B-ResNet}.
\section{Conclusion}
\label{sec:conclusion}
Bringing DGA detection closer to practice, we proposed ResNet-based classifiers for the DGA binary and multiclass classification task and compared their performance to various state-of-the-art classifiers in a unified setting using the same real-world NX-traffic. B-ResNet achieves accuracies comparable to the other classifiers while it exceeds them in the extraction of complex features. Our M-ResNet.MI model outperforms the other classifiers while it reduces the required training time by over 30\% compared to the next best classifier. We demonstrated that B-ResNet is remarkably time-robust and generalizes well to new environments, which allows for the provisioning of classification as a service. B-ResNet, as well as M-ResNet.MI, are able to perform live detection in large networks and are able to detect both, new DGAs as well as new seeds of known DGAs. In particular, in our one-month real-world application test we could discover 8 new DGA-related groups of AGDs, 6 of which appear to originate from yet unknown DGAs. In our explainability analysis, we \mbox{highlighted some of the self-learned features} used by the deep learning based approaches. The classifiers are robust against two recently published adversarial attacks when trained on NX-traffic (\cite{peck_charbot_2019,spooren_detection_2019}). This unfolds yet another advantage of using NX-traffic as training data. In future work, we plan to investigate the effectiveness of more complex attacks (e.g.\cite{anderson_deepdga_2016,sidi_maskdga_2019}) as well.

\begin{acks}
The authors would like to thank Daniel Plohmann for granting us access to DGArchive as well as Jens Hektor and Siemens for providing NXD data. This project has received funding from the European Union's Horizon 2020 research and innovation programme under grant agreement No 833418. Simulations were performed with computing resources granted by RWTH Aachen University under project rwth0438.
\end{acks}

\bibliographystyle{ACM-Reference-Format}
\bibliography{bibliography}


\end{document}